\documentclass[preprint,prl,showpacs,
amsmath,amssymb,floatfix]{revtex4}
\usepackage{graphicx}
\usepackage{dcolumn}
\usepackage{bm}

\begin{document}
\title{Transverse momentum distribution of vector
  mesons produced in ultraperipheral relativistic heavy ion collisions}
\author{Kai Hencken}
\affiliation{
Institut f\"ur Physik, Universit\"at Basel, 4056 Basel, Switzerland}
\author{Gerhard Baur}
\affiliation{
Institut f\"ur Kernphysik, Forschungszentrum J\"ulich, Postfach 1913,
52425 J\"ulich, Germany}
\author{Dirk Trautmann}
\affiliation{
Institut f\"ur Physik, Universit\"at Basel, 4056~Basel, Switzerland}

\date{\today}

\begin{abstract}
We study the transverse momentum distribution of vector mesons
produced in ultraperipheral relativistic heavy ion collisions (UPCs).
In UPCs there is no strong interaction between the nuclei and the
vector mesons are produced in photon-nucleus collisions where the 
(quasireal) photon is emitted from the other nucleus.
Exchanging the role of both ions leads to interference effects.  
A detailed study of the transverse momentum distribution 
which is determined by the transverse momentum of the emitted photon, 
the production process on the target and the interference
effect is done. We study the total unrestricted cross section
and those, where an additional electromagnetic
excitation of one or both of the ions takes place in addition to the 
vector meson production, in the latter case small impact parameters
are emphasized.
\end{abstract}


\maketitle
Due to the strong electromagnetic fields surrounding the heavy ions in
relativistic collisions, RHIC and LHC can be seen as a factory of
quasireal photons of high energies. One of the interesting
photonuclear processes studied in these ``ultraperipheral collisions'' (UPC) 
is the coherent production of vector mesons, in particular $\rho^0$,
which has been measured recently at RHIC \cite{Adler:2002sc,Klein:2004kq}. 
The coherent production was identified through the transverse momentum
distribution of the meson, which is enhanced for values of the
transverse momentum 
 $v_\perp\lesssim 1/R$ where $R$ denotes the nuclear radius. 
We give a careful theoretical study of the process
\begin{equation}
A + A \rightarrow A^{(*)} + A^{(*)} + V
\label{eq:genprocess}
\end{equation}
with (``$A^*$'') and without (``$A$'') an electromagnetic (GDR)
excitation of either one or both ions.
This is of interest for the analysis of the RHIC
experiments, as well as, for future experiments at LHC
where also other vector mesons like
($J/\Psi$ and even $\Upsilon$) can be studied
\cite{Frankfurt:2001db,Frankfurt:2003qy,Bertulani:2005ru}.
While the theory of UPC is generally in a good 
shape \cite{Baur:2001jj,Baltz:2002pp,BaurHT98,Krauss:1997vr}, 
the specific question of the {\em transverse momentum distribution}
has been paid attention to only in less rigor
\cite{Baltz:2002pp,Klein:1999gv,Klein:1999qj}.

Heavy ion scattering offers a unique possibility
to study an important interference effect \cite{Klein:1999gv}.
As will be shown below, the transverse momentum distribution is 
very sensitive to this effect. 
It is the purpose of this letter to give a careful study
of this transverse momentum distribution.
\begin{figure}[tbh]
  \centering
     \includegraphics[width=2.5cm]{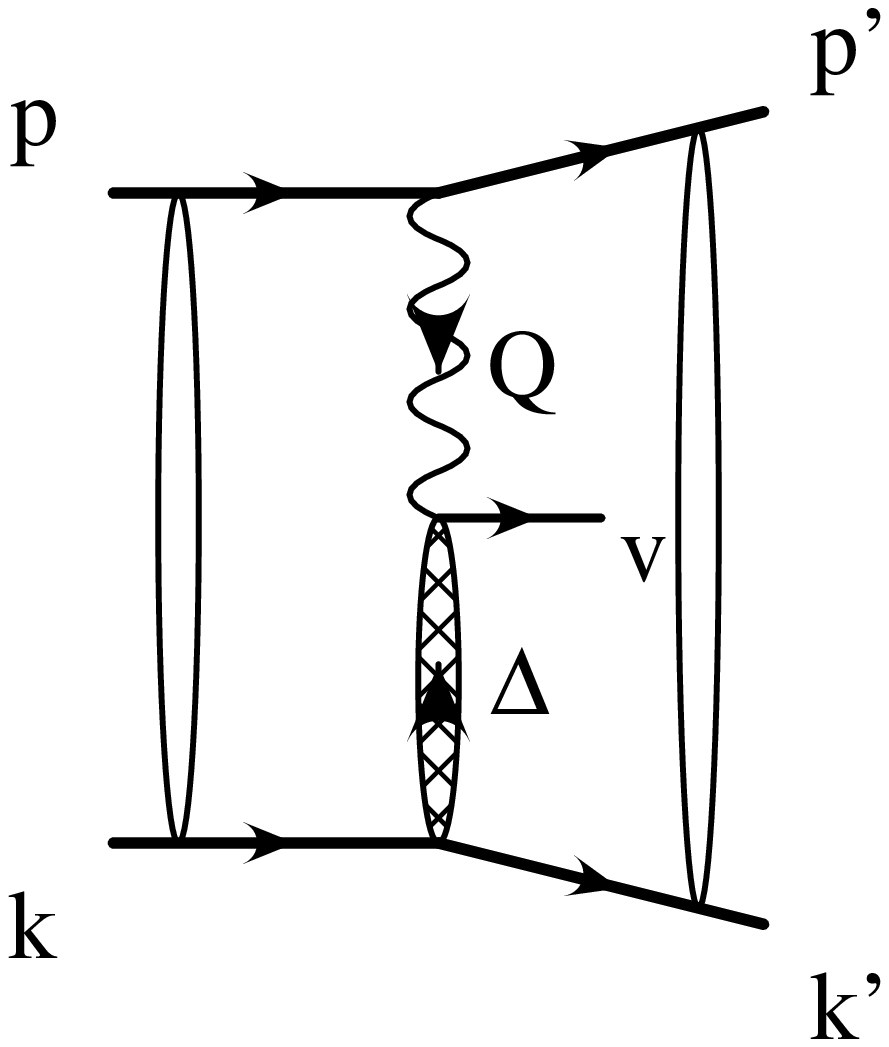}(a)
     \includegraphics[width=2.5cm]{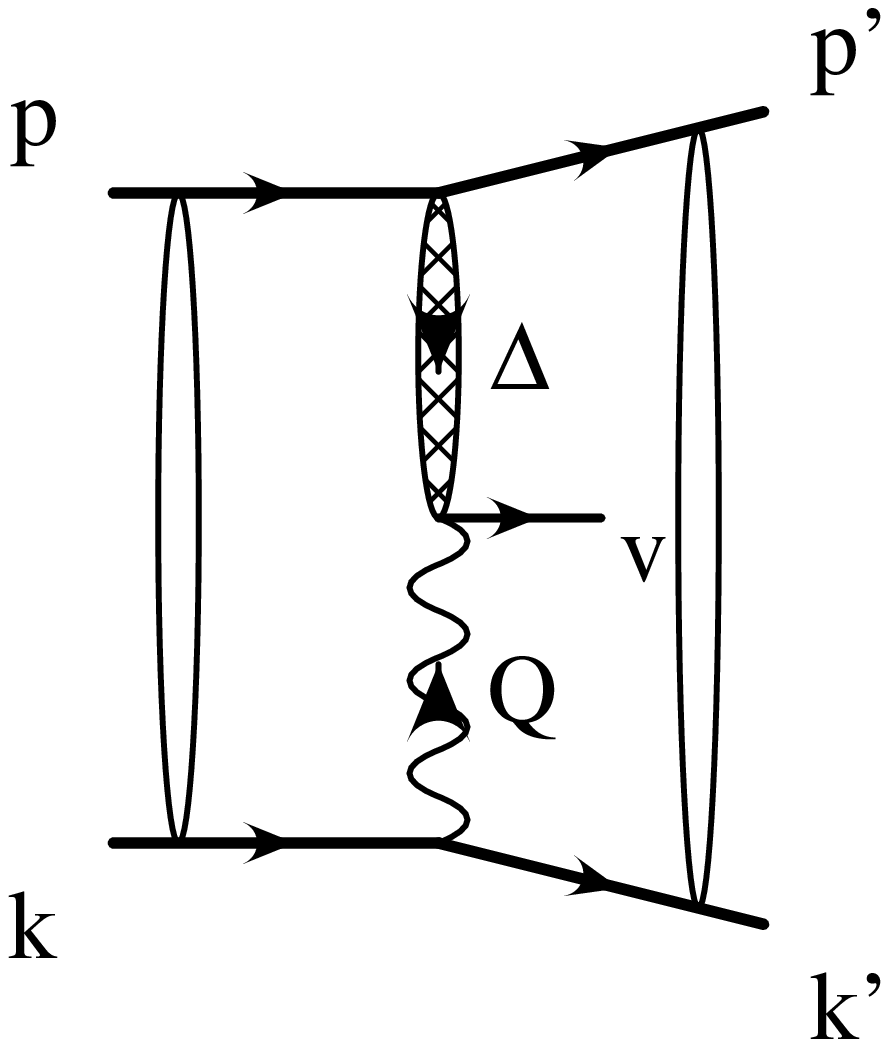}(b)
\caption{A schematic Feynman diagram for the vector meson production
in ultraperipheral heavy ion collisions (a). The corresponding exchange
diagram is also shown (b).}
\label{fig:AAV}
\end{figure}

The  kinematics of the process given in Eq.~(\ref{eq:genprocess}) is
denoted by (see Fig.~\ref{fig:AAV})
\begin{equation}
p + k \rightarrow p' + k' + v.
\end{equation}
Due to the additional elastic photon exchanges
which are schematically denoted by the open blobs in
Fig.~\ref{fig:AAV} the momenta  $Q$ and $\Delta$ are 
not related to the asymptotic momenta  by $Q=p-p'$ and $\Delta=k-k'$
as, e.g., in the $pp$ case (without rescattering) \cite{Khoze:2002dc}. 
For small transverse momenta the longitudinal components of the photon
momentum and the momentum transfer from the vector meson production
(``Pomeron momentum'') 
are given in the c.m. system by
the mass $m_V$ and the rapidity $Y$ of the produced
meson  as 
\begin{equation}
Q_0 = \beta Q_z = \frac{m_V}{2} e^Y, \ 
\Delta_0 = - \beta \Delta_z =  \frac{m_V}{2} e^{-Y}.
\label{eq:longkinematics}
\end{equation}
The momenta $p$ and $k$, see Fig.~\ref{fig:AAV}(a),
are given by $p=m_A u_+$ (ion 1) and $k=m_A u_-$
(ion 2), where $m_A$ is the ion mass and $u_{\pm}=\gamma (1,0,0,\pm \beta)$.
In the exchanged process, see Fig.~\ref{fig:AAV}(b),
the photon is emitted from ion 2, the ``Pomeron'' from ion 1.

Due to the large value of the Coulomb 
parameter $\eta=\frac{Z_1 Z_2e^2}{\hbar v}$ we can use the semiclassical
approximation \cite{Klein:2004kq,Klein:2003qc,Baur:2003ar}.
We also show how this can be derived from eikonal/Glauber theory,
more details will be given in a forthcoming publication
\cite{HenckenTBP05}. 
Using a simple model for the meson production process
we are able to give analytical results.
Implications for the current
experiments at STAR/RHIC and for future experiments at the LHC
\cite{Klein:1999gv,Frankfurt:2003qy} are discussed.
In the semiclassical approximation the two ions move along a
straight line and the process is described by an impact parameter
dependent amplitude $a(\vec b,\vec v_\perp, Y)$. In contrast to the 
momentum of the vector meson 
the momenta of the outgoing ions are not detected and 
the differential cross section is given by
\begin{equation}
\frac{d^3\sigma}{d^2v_\perp dY} = 
\frac{1}{2 (2\pi)^3} 
\sum_{e_V} \int d^2b |a_{fi}(\vec b,\vec v_\perp,Y)|^2.
\end{equation}
The integration over the impact parameter $\vec b$
corresponds to an integration over the unobserved momenta $k'$
and $p'$ of the scattered ions. The different processes which can occur 
according to Eq.~(\ref{eq:genprocess}) factorise \cite{Baur:2003ar}: 
\begin{equation}
a_{fi}(\vec b,\vec v_\perp,Y) = a_{nucl}(\vec b) \  a_{1}(\vec b) \
a_{2}(\vec b) 
\ a_{V}(\vec b,\vec v_\perp,Y).
\label{eq:semiclass1}
\end{equation}
The strong absorption due to the interaction of the ions for $b<2R$
is given by $a_{nucl}(\vec b)\approx \Theta(b-2R)$ with the nuclear radius
$R = 7$fm.
$a_V(\vec b,\vec v_\perp,Y)$ describes the vector meson production.
Additional electromagnetic excitation amplitudes of ion $1$ and/or $2$ are
denoted by $a_{1}(\vec b)$ and $a_{2}(\vec b)$. What is chosen for
them depends on the additional triggering condition used in the experiment.
The cross section can be written as
\begin{eqnarray}
\frac{d^3\sigma}{d^2v dY} 
&=& 
\frac{1}{2 (2\pi)^3} \int_{2R}^\infty d^2b f_{ij}(b)
\sum_{e_V} |a_{V}(\vec b,\vec v,Y)|^2.
\label{eq:semiclass2}
\end{eqnarray}
where $f_{ij}(b)$ takes the triggering condition of the measurement
into account. In the case
where the process is observed without any further condition imposed on
the excitation of the ion(s), we have $f_{00}(b)=1$. If the vector
meson production together with the electromagnetic excitation is
measured one uses either $f_{10}(b)=P_1(b)=|a_1(b)|^2$ or 
$f_{01}(b)=P_2(b)=|a_2(b)|^2$ if
the excitation of one of the ions, $f_{11}(b)=P_1(b)P_2(b)$ if the
mutual excitation of both ions is triggered on.

The electromagnetic excitation probabilities 
are given by $P_{i}(b) = \frac{S}{b^2}$
with $S \approx 5.45 \times
10^{-5} Z^3 N A^{-2/3} \mbox{fm}^2$ \cite{BaurB88}.
This equation is valid for
$2R<b<b_{max}=\frac{2\gamma^2-1}{E_{GDR}}$, but 
this cutoff can be safely neglected here, since the vector meson production
probability falls off at least as fast as  $1/b^2$.

In the semiclassical treatment  the amplitude of an electromagnetic
process can be written as \cite{BaurB93,VidovicGB93}
\begin{equation}
  a(\vec b) = \int \frac{d^4Q}{(2\pi)^4} A_{ext}^\mu (\vec b,Q) 
  J_{\mu}(Q),
\label{eq:ascl}
\end{equation}
where
\begin{equation}
A^\mu_{ext}(b,Q) =  2 \pi Z e Q_\perp^\mu \delta(Q u_+)  \frac{\gamma}{Q_0} 
\frac{F(Q^2)}{Q^2} \exp(- i \vec Q_\perp \vec b)
\label{eq:lienard}
\end{equation}
is the Li\`enard-Wiechert potential. A gauge transformation has
been made so that the field is to a good approximation transversal. 

For the elastic form factor $F(Q^2)$ we choose 
$F(Q^2) = \exp(Q^2 R_\gamma^2)$, with
$R_\gamma\approx\sqrt{<r^2>/6}\approx 2.2$fm.
Alternatively we can set $F(Q^2)=1$, that is $R_\gamma=0$, as the
electric field outside a spherically symmetric charge distribution is
the same as that of a corresponding point charge. 
We find numerically the effect of a finite
$R_\gamma$ to be rather small, justifying this assumption. We still keep
it for completeness in the following equations.

In order to describe the meson production we need an expression for
the electromagnetic current $J(A \rightarrow A+V)$. 
This current can be found by using the vector dominance model (VDM),
which relates this current to the elastic scattering amplitude 
$V+A \rightarrow V+A$ as
\begin{equation}
J^\mu(Q) = e^\mu_V C_V f_{el}(\Delta_\perp,Y)
\label{eq:jqd}
\end{equation}
with $\Delta_\perp = v_\perp - Q_\perp$. 
$e_V$ is the polarisation of the outgoing vector meson,
which by assuming $s$-channel helicity conservation (SCHC) is
identical to the one of the incoming photon, see
\cite{Crittenden:1997yz} 
for details. $C_V$ describes the vector meson content of the photon,
see e.g. \cite{Bertulani:2005ru}.

In the following we choose the elastic vector meson scattering
amplitude as
\begin{equation}
f_{el}(\Delta_\perp,Y) = f_0(Y) \exp(-\Delta_\perp^2 R_V^2)
\delta(\Delta u_-) (v u_-)
\label{eq:fel}
\end{equation}
with $R_V\approx2.2$fm to reproduce the slope of the angular distribution.
This form agrees also with the one proposed in 
\cite{Kopeliovich:2001dz}.
It has been mainly chosen for ease of calculations, 
whereas the formalism given
can be extended to more realistic forms, e.g., based on a full eikonal
description. A simple extension is possible by using a sum of
Gaussians for $f_{el}$.

The imaginary part of $f_0$ can be related to the total cross section
for vector meson scattering on nuclei through the optical theorem. For
the energies of RHIC and LHC the real part is small (of the order of
10\%). While for $Y\not =0$ (and asymmetric collisions) there is a 
sensitivity to the phase of $f_0$,
for rapidity $Y=0$ we only need the absolute value of $|C_V f_0|$.
This we choose in order to reproduce the cross section given in
\cite{Frankfurt:2002wc,Frankfurt:2001db} as 
$d\sigma/dY_{\rho}(\mbox{RHIC})=70 mb$
and $d\sigma/dY_{J/\Psi}(\mbox{LHC})=0.75 mb$.

Using these expressions we get for the amplitude:
\begin{eqnarray}
&&a_V(\vec b,\vec v_\perp,Y) =
\frac{Z e C_V f_0}{(2 \pi)^3} \exp(-Q_l^2 R_\gamma^2)
\nonumber\\ &&\int d^2Q_\perp
(\vec Q_\perp \vec e_V)
\frac{\exp(-Q_\perp^2 R_\gamma^2)}
{Q_\perp^2+Q_l^2}  \nonumber\\
&& \exp(-i \vec Q_\perp \vec b) \exp(-R_V^2 (\vec v_\perp-\vec Q_\perp)^2)
\label{eq:aV}
\end{eqnarray}
with $Q_l^2=Q_z^2-Q_0^2=(\frac{Q_0}{\beta \gamma})^2$.
Let us first make a short qualitative discussion: if we 
would neglect $\vec Q_\perp$ in $\exp(-R_V^2 (\vec v_\perp-\vec
Q_\perp)^2)$, the dependence
of $|a_V|^2$ and $d^3\sigma/d^2v_\perp dY$ on $v_\perp$ would
be of the form $\exp(-2 R_V^2 v_\perp^2)$, which is due to $J$ alone. 
This coincides with the result for an incident photon of zero
transverse momentum. The effect of the finite $Q_\perp$ distribution
of the photon is to broaden this distribution. As the width of the
$Q_\perp$ distribution depends on $b$ via $\exp(- i \vec Q_\perp \vec
b)$, the effect of this broadening will depend on $b$. This effect is
largest for small $b$, as the perpendicular momentum distribution of
the photon is of the order $1/b$.

An analytic approximation for $a_V$ can be found in the region of
small $b$, that is if $b< 1/Q_l$ and $Q_l$ can be neglected in
the photon propagator in Eq.~(\ref{eq:aV}), 
which corresponds to the sudden limit. One gets (see
\cite{HenckenTBP05} for details):
\begin{eqnarray}
a_{V}(\vec b, \vec v_\perp, Y) &\approx&
\frac{Z e C_V f_0 i}{(2\pi)^2}
\frac{(\vec b + 2 i \vec v_{\perp} R_V^2) \vec e_V }
{(\vec b +  2 i \vec v_\perp R_V^2)^2} \nonumber\\
&&\exp(-Q_l^2 R_\gamma^2) \exp(-v_\perp^2 R_V^2) \nonumber\\
&&\left[ \exp\left(\frac{-(\vec b+2 i \vec v_\perp R_V^2)^2}{4
      (R_V^2+R_\gamma^2)} \right)-1 \right]. 
\label{eq:analytic}
\end{eqnarray}

The same final state can be obtained by exchanging the roles of both
ions, see Figs.~\ref{fig:AAV}(a) and~(b).
The corresponding amplitude $a_{V}^X(\vec b,\vec v_\perp,Y)$ 
where the photon is emitted from ion 2 and the ``Pomeron'' from ion 1 
is given by
\begin{equation}
a_{V}^X(\vec b,\vec v_\perp, Y) = \int \frac{d^4Q}{(2\pi)^2} 
A_{ext}^\mu (0,Q) 
J^X_{\mu}(Q),
\end{equation}
where the impact parameter for $A_{ext}$ is now $\vec b=0$
and $u_+$ is replaced by $u_-$.
The electromagnetic current $J^X$ is now for vector meson
production on an ion at position $\vec b$. One finds
\begin{eqnarray}
J^X_{\mu}(Q) &=& J_{\mu}(Q) \exp(-i \vec Q_\perp \vec b)
\nonumber \\
&=&  J_{\mu}(Q) \exp(-i \vec v_\perp \vec b+i \vec
\Delta_{\perp} \vec b).
\end{eqnarray}

We find that the exchange amplitude is of the form:
\begin{equation}
a_V^X(\vec b, \vec v_\perp,Y) = a_V(-\vec b,\vec v_\perp, -Y) 
\exp(-i \vec v_\perp \vec b).
\end{equation}
This has a simple interpretation: In order to exchange the role of the
two ions, $Y$ is replaced by $-Y$, the direction of $\vec b$ needs to 
be reversed and in addition the origin needs to be shifted by $\vec
b$, leading to the extra phase
$\exp(- i \vec v_\perp \vec b)$. This relation was also used in
\cite{Nystrand:2002pd,Klein:1999gv}. With $a_V$ from
Eq.~(\ref{eq:aV}) we finally get
\begin{equation}
a_{V}^{tot}(\vec b,\vec v_\perp,Y) = 
a_{V}(\vec b,\vec v_\perp,Y) + 
e^{- i\vec v_\perp \vec b} a_{V}(-\vec b,\vec v_\perp,-Y).
\end{equation}

The analytic expression in Eq.~\ref{eq:analytic} 
allows us to discuss some properties of the transverse
momentum distribution of the process: In the limit $v_\perp R_V^2\ll b$ one
has $a_V \sim \vec b \vec e_V$ and $a_V^X=-a_V$, i.e., the
amplitudes have a relative sign of $-1$, leading to destructive
interference at small $b$.
In the other limit $v_\perp R_V^2 \gg b$ one has
$a_V^X=a_V$, i.e., the same relative sign, but $a_V$ and $a_V^X$ are
smaller than in the first case due to the last exponential in
Eq.~(\ref{eq:analytic}). The transverse momentum distribution is
therefore more complex than treated in \cite{Klein:1999gv,Nystrand:2002pd}.

We can also derive the results starting from the eikonal or Glauber approach
to multiphoton processes in UPC collisions, see \cite{Baur:2003ar}. In
this case the scattering amplitude is given by
\begin{equation}
f_{fi,\mbox{Glauber}}(\vec K)= \frac{i\pi}{k} \int d^2b \exp(i \vec K \vec b) 
\left< f \right| \exp(i\chi(\vec b)) \left| i \right>,
\end{equation}
where $\vec K=\vec k'-\vec k$ is the total momentum transfer to the
``target'' nucleus. The eikonal $\chi(b)$ takes care of all the different
elastic and inelastic processes. In our case we have
\begin{equation}
\chi(\vec b)= \chi_{nuc}(b) + \chi_{C}(b) + \chi_{1}(b) + \chi_{2}(b)
+ \chi_{V}(\vec b).
\end{equation}
The term $\chi_{nuc}(b)$ describes the effect coming from the nuclear
interaction between the two ions. It can be approximated by $\exp(i
\chi_{nuc}(b)) \approx \Theta(b-2R)$. The term
$\chi_{C}\approx \exp(2 i \eta \log(kb))$ describes the elastic
Coulomb scattering. The last three terms describe the additional 
electromagnetic interactions: the possible excitation of the first and second
nucleus and the vector meson production.
The eikonal phase for the vector meson production process $\chi_V(b)$ can
usually be treated in lowest order by expanding the exponential. 
The second order term would describe double $\rho$ production, which
is still sizeable, see \cite{Klein:1999qj}. 
Bracketing with the initial and final states we get
\begin{eqnarray}
&&<f| \exp(i \chi(\vec b)) |i> \approx i
\exp(i \chi_{nuc}(b)) \exp(+i \chi_{C}(b)) \nonumber\\ &&
<f_1 |\exp(i \chi_1(b))|i_1> <f_2|\exp(i \chi_2(b))|i_2> \nonumber\\&&
<V,i_2|\chi_{V}(\vec b)|i_2>   
\end{eqnarray}
with $|i>=|i_1,i_2>$ the initial (ground) states of the two ions and
$|f>=|f_1,f_2>|V>$ the final states of the ions and the meson.
In order for this process
to factorise, we made the reasonable assumption, that the vector
meson production on the excited nucleus is the same as the one on the
nucleus in the ground state:
\begin{equation}
<V,i_2|\chi_{V}(b)|i_2> \approx <V,f_2|\chi_{V}(b)|f_2>.
\end{equation}

There is a correspondence of these terms to the different
semiclassical amplitudes $a_{fi}(b)$ in Eq.~(\ref{eq:semiclass1}),
which was also explored in \cite{Baur1991}. The $\chi$ is given by
\begin{equation}
\chi(\vec b) = -1 \int \frac{d^4Q}{(2\pi)^4}
A^{\mu}_{\mbox{eik}}(\vec b,z,Q) \hat J_\mu(Q)
\end{equation}
with $A^{\mu}_{\mbox{eik}}$ as given in Eq.~(\ref{eq:lienard}) with
  $\delta(Q u_+)$ replaced with $\delta(\gamma (Q_0-Q_z))$,
which corresponds to the expression of the semiclassical amplitude $a_V$ in
Eq.~(\ref{eq:ascl}) in the sudden limit. 

The major difference between the two approaches is the presence of the
Coulomb eikonal $\chi_{C}(b)$. 
For $\eta\gg 1$, $\chi_{C}(b)$ is a rapidly varying function and
one can evaluate the Glauber expression by means of the well known
saddle point approximation. One obtains the relation between the classical
impact parameter and the momentum transfer: $b=2\eta/K$.
As the momentum transfer to the ion is not measured in our case,
one calculates ``inclusive'' cross sections by integrating over $K$. 
This gives the same result as in the semiclassical case of 
Eq.~(\ref{eq:semiclass1}) in the sudden limit ($Q_l=0$),
as the Coulomb phase is purely imaginary and the integration over $K$ leads,
(see Eq.~(10)-(12) of \cite{Baur1991}) 
to a $\delta$-function for the two impact parameters in 
$|f_{fi, \mbox{Glauber}}|^2$.

\begin{figure}[tbh]
  \centering
     \includegraphics[height=0.31\hsize]{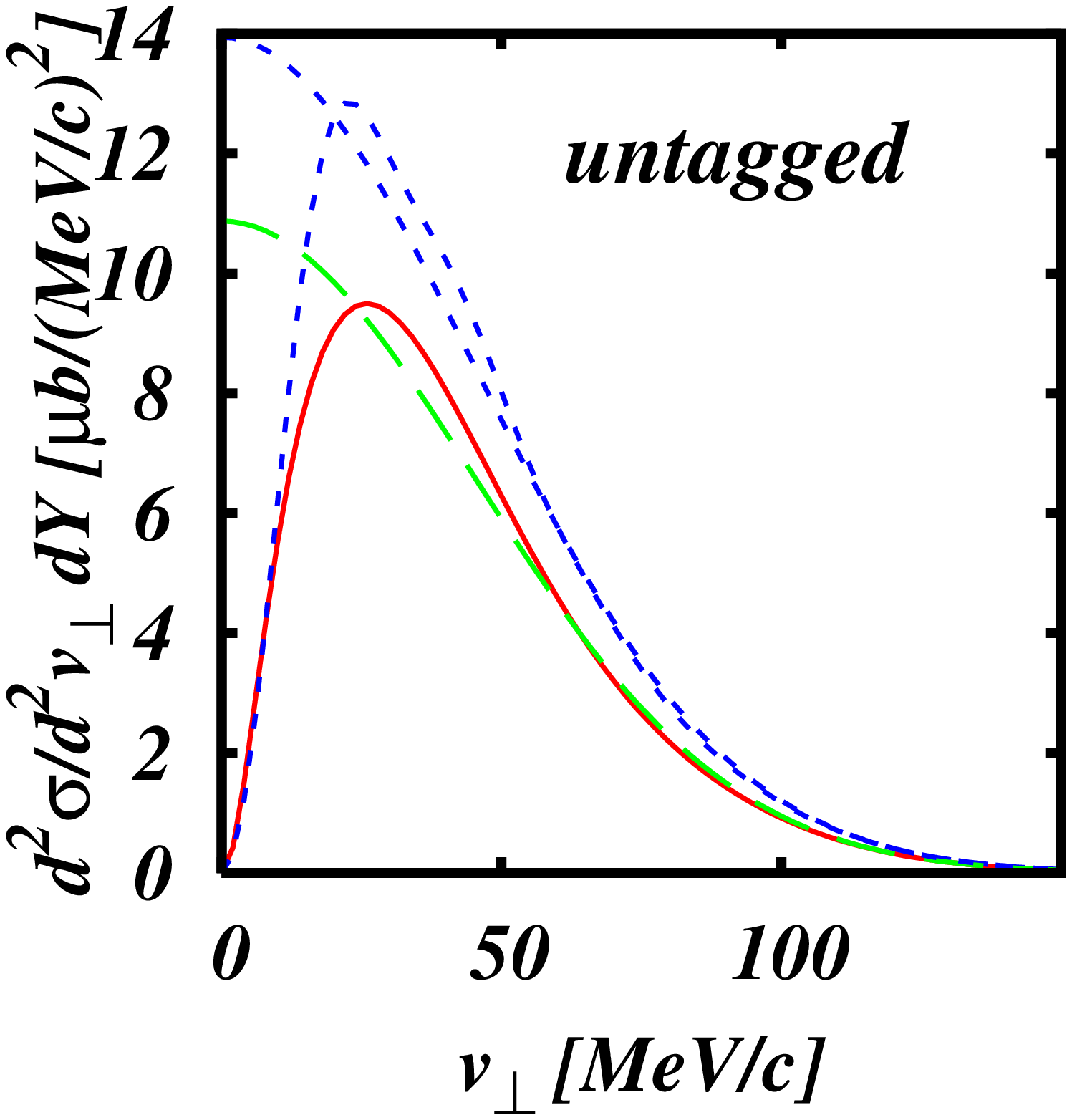}
     \includegraphics[height=0.31\hsize]{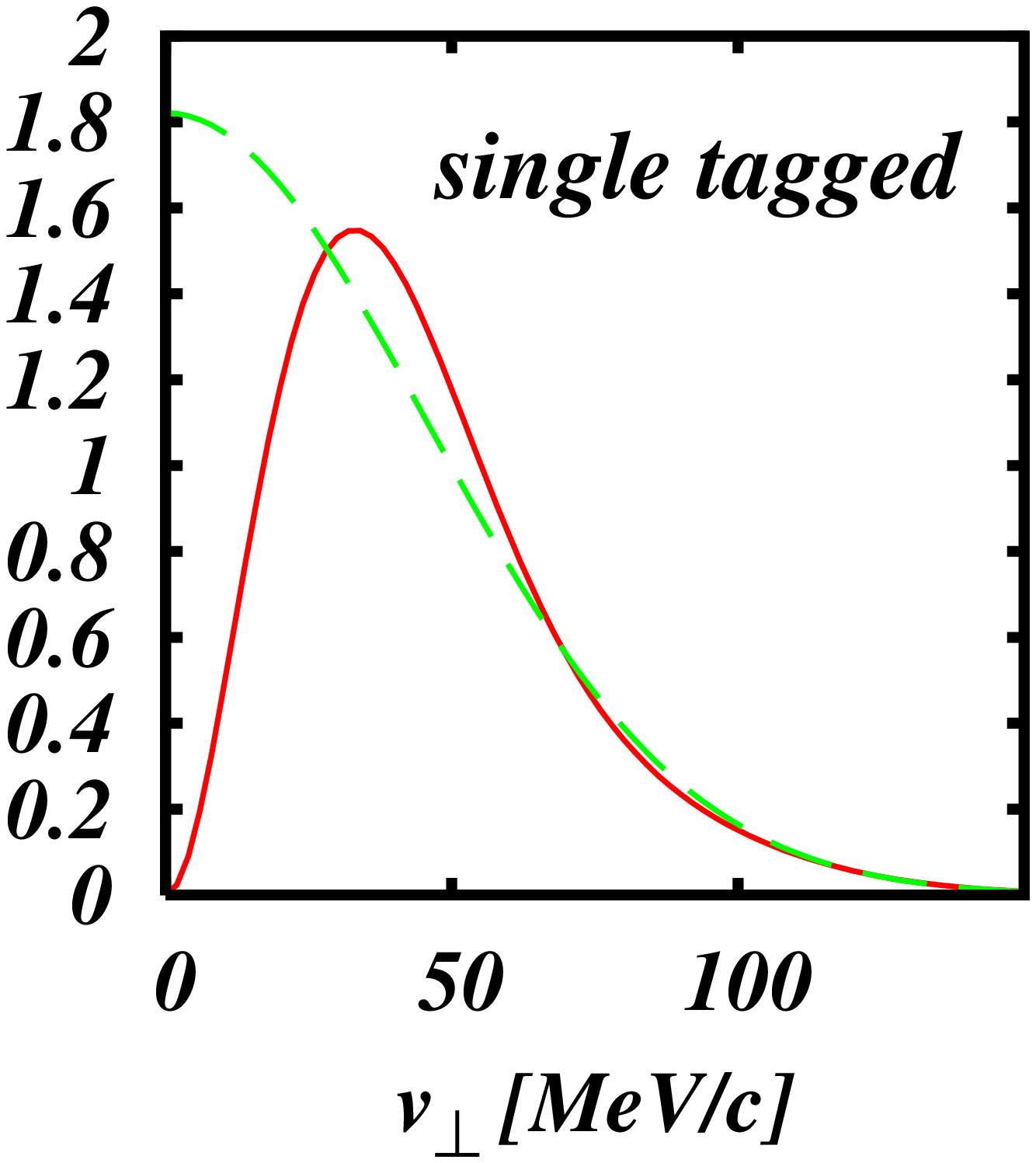}
     \includegraphics[height=0.31\hsize]{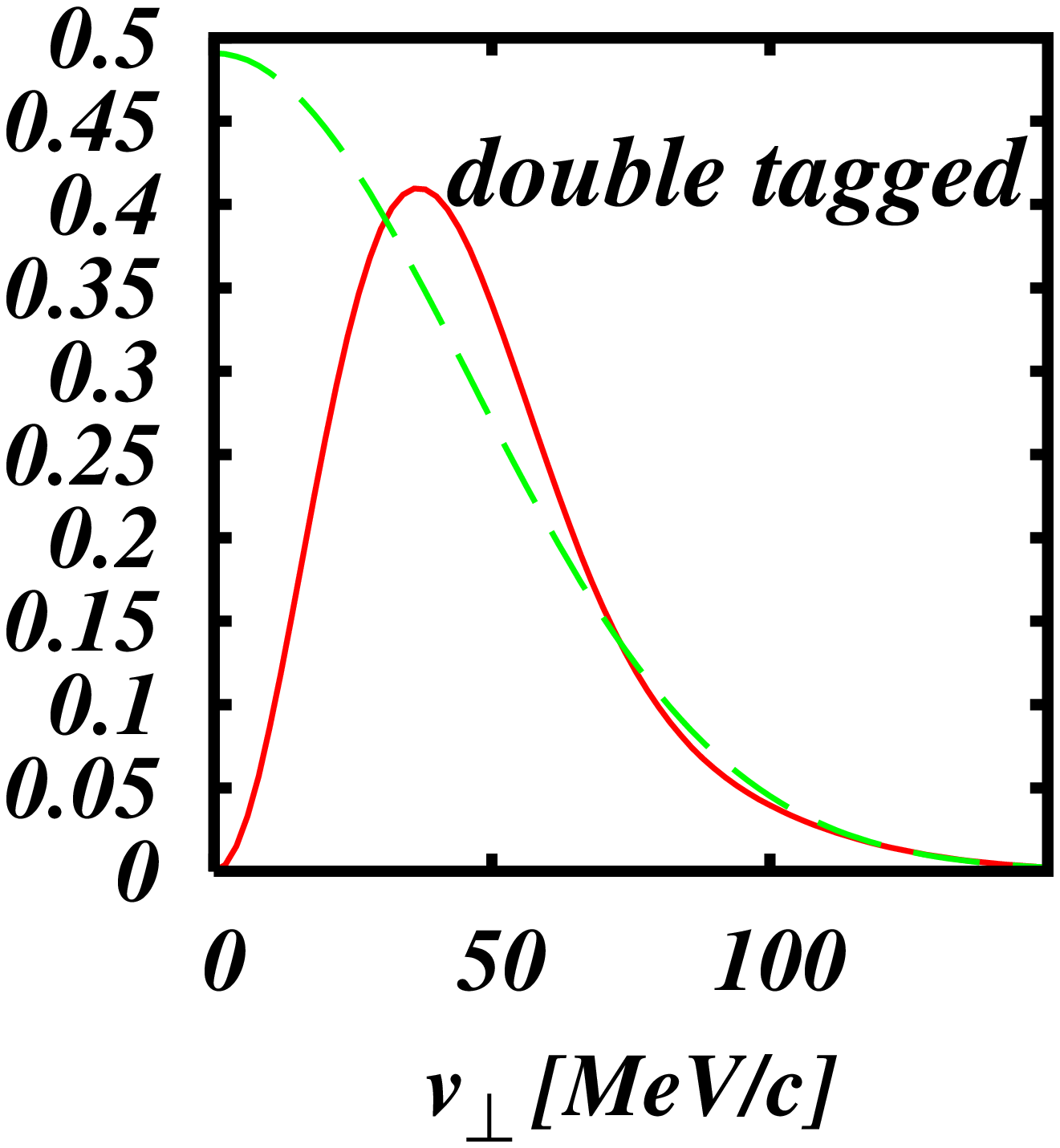}
\caption{The differential cross section $d\sigma/d^2v_\perp dY$ is
shown for $\rho^0$ production at $Y=0$ at RHIC (Au-Au collisions with
$\gamma=108$). The solid line is the result including the
interference, the dashed line the result from an incoherent adding of
the two processes. The results for the three different tagging cases
are given in (a), (b) and (c). 
The approximate result is shown as dotted line only in the first
(untagged) case. In the other two case it cannot be distinguished from
the full calculation.}
\label{fig:norm1}
\end{figure}

\begin{figure}[tbh]
  \centering
     \includegraphics[height=0.31\hsize]{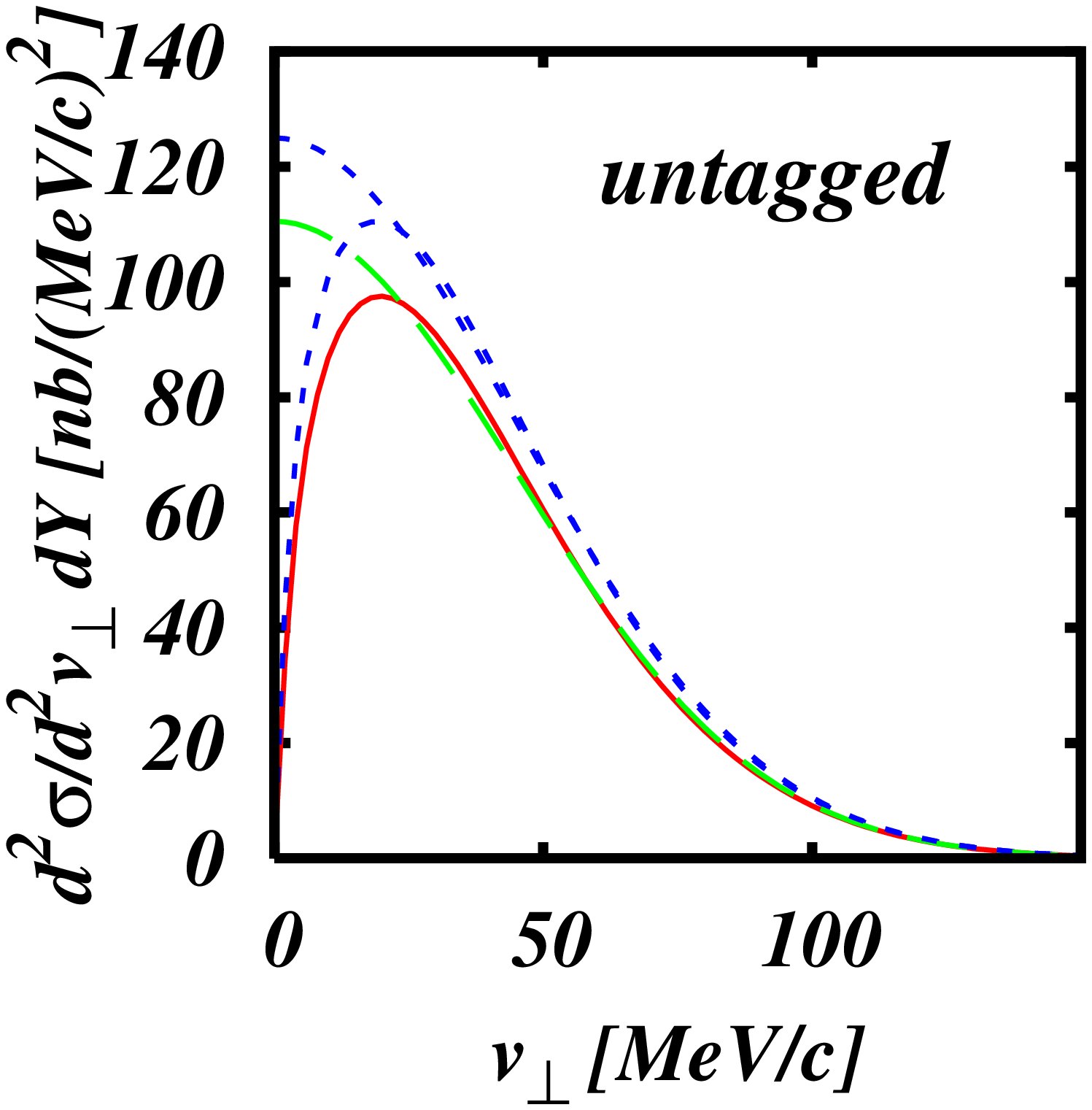}
     \includegraphics[height=0.31\hsize]{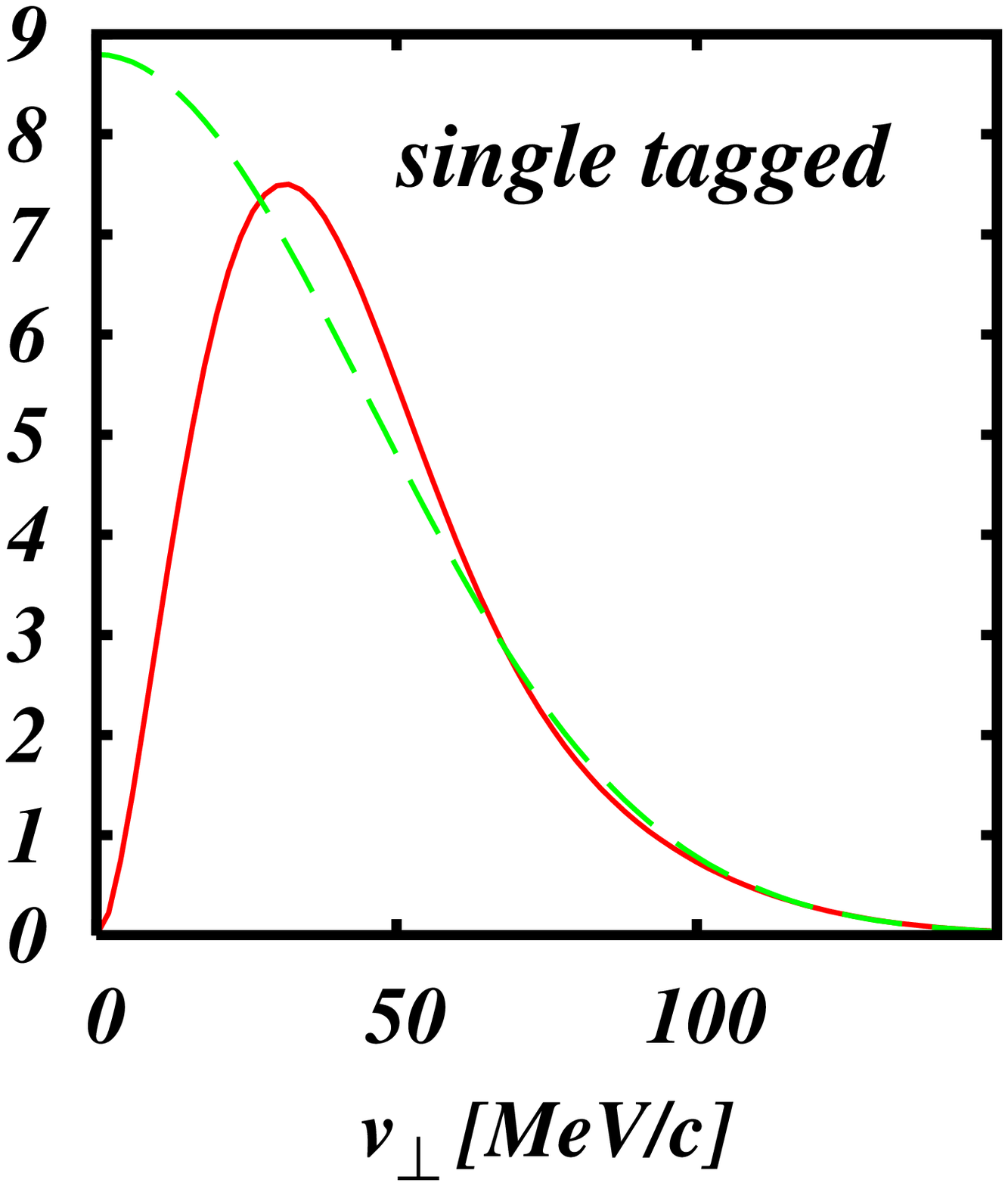}
     \includegraphics[height=0.31\hsize]{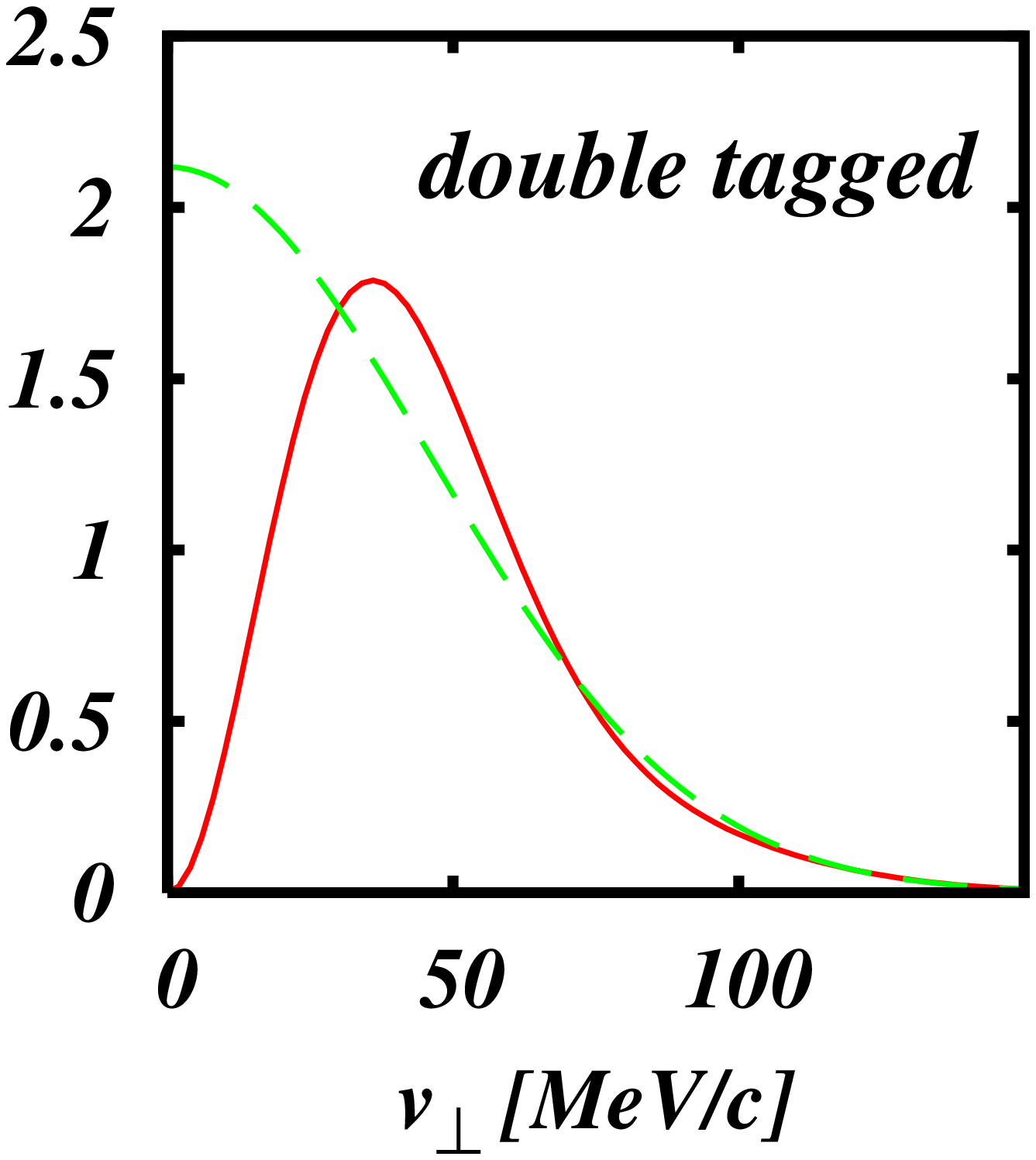}
\caption{The differential cross section $d\sigma/d^2v_\perp dY$ is
shown for $J/\Psi$ production at $Y=0$ at LHC (Pb-Pb ions with
$\gamma=3000$). The lines are the same as in Fig.~\protect{\ref{fig:norm1}}.}
\label{fig:norm2}
\end{figure}

We use both the exact expression Eqs.~(\ref{eq:semiclass2})
and~(\ref{eq:aV}), as well as, the approximate analytical
result Eq.~(\ref{eq:analytic}) to calculate results for the case 
studied in \cite{Klein:1999gv}. They are shown in Fig.~\ref{fig:norm1}.
The analytic result is too large in the untagged case, but its shape
agrees quite nicely with the full calculation. The effect of
tagging for small $b$ is a shift of the maximum of the curve to larger
values of $v_\perp$ and a more pronounced interference structure.
In Fig.~\ref{fig:norm2} we also show the similar results for $J/\Psi$
production at the LHC. 

Let us summarize our findings: We have put the transverse momentum
distribution on a firm theoretical basis starting our derivation from
the semiclassical approximation or alternatively from Glauber theory. 
The meson transverse momentum distribution was derived as a function
of $b$ and an analytic expression was given. The interference
phenomenon was derived within this model.
As the main outcome we find in this letter that for a good understanding 
of the interference phenomenon a careful study of the transverse
momentum distribution is essential. Whereas formally the
results look similar to the one given in 
\cite{Klein:1999gv,Klein:2004kq}, 
differences appear both in the transverse momentum distribution as a
function of $b$ and in the form of the interference. This leads to a
more complex result in the intermediate $v_\perp$ region. This will
also be true in the case, where one is moving away from $Y=0$.
Our findings are important in analyzing the
experimental data of STAR and also PHENIX. Results have also been
given for future LHC measurements, which would be even more
interesting for $J/\Psi$ or even $\Upsilon$ production.

\newpage
Figure 2:\\
     \includegraphics[height=0.5\hsize]{Figs/rho1.eps}
     \includegraphics[height=0.5\hsize]{Figs/rho2.eps}
     \includegraphics[height=0.5\hsize]{Figs/rho3.eps}
\newpage
Figure 3:\\
     \includegraphics[height=0.5\hsize]{Figs/jpsi1.eps}
     \includegraphics[height=0.5\hsize]{Figs/jpsi2.eps}
     \includegraphics[height=0.5\hsize]{Figs/jpsi3.eps}

\begin{thebibliography}{10}
\bibitem{Adler:2002sc}
  C.~Adler {\it et al.}  [STAR Collaboration],
  Phys.\ Rev.\ Lett.\  {\bf 89}, 272302 (2002)
  [arXiv:nucl-ex/0206004].

\bibitem{Klein:2004kq}
  S.~R.~Klein  [STAR Collaboration],
  arXiv:nucl-ex/0402007.

\bibitem{Frankfurt:2001db}
  L.~Frankfurt, M.~Strikman and M.~Zhalov,
  Phys.\ Lett.\ B {\bf 540}, 220 (2002)
  [arXiv:hep-ph/0111221].

\bibitem{Frankfurt:2003qy}
  L.~Frankfurt, V.~Guzey, M.~Strikman and M.~Zhalov,
  JHEP {\bf 0308}, 043 (2003)
  [arXiv:hep-ph/0304218].

\bibitem{Bertulani:2005ru}
  C.~A.~Bertulani, S.~R.~Klein and J.~Nystrand,
  arXiv:nucl-ex/0502005.

\bibitem{Baur:2001jj}
  G.~Baur, K.~Hencken, D.~Trautmann, S.~Sadovsky and Y.~Kharlov,
  Phys.\ Rept.\  {\bf 364}, 359 (2002)
  [arXiv:hep-ph/0112211].

\bibitem{Baltz:2002pp}
  A.~J.~Baltz, S.~R.~Klein and J.~Nystrand,
  Phys.\ Rev.\ Lett.\  {\bf 89}, 012301 (2002)
  [arXiv:nucl-th/0205031].

\bibitem{BaurHT98}
G. Baur, K. Hencken, and D. Trautmann, Topical Review, J. Phys. G {\bf 24},
  1657  (1998).

\bibitem{Krauss:1997vr}
  F.~Krauss, M.~Greiner and G.~Soff,
  Prog.\ Part.\ Nucl.\ Phys.\  {\bf 39}, 503 (1997).

\bibitem{Klein:1999gv}
  S.~R.~Klein and J.~Nystrand,
  Phys.\ Rev.\ Lett.\  {\bf 84}, 2330 (2000)
  [arXiv:hep-ph/9909237].

\bibitem{Klein:1999qj}
  S.~Klein and J.~Nystrand,
  Phys.\ Rev.\ C {\bf 60}, 014903 (1999)
  [arXiv:hep-ph/9902259].

\bibitem{Khoze:2002dc}
  V.~A.~Khoze, A.~D.~Martin and M.~G.~Ryskin,
  Eur.\ Phys.\ J.\ C {\bf 24}, 459 (2002)
  [arXiv:hep-ph/0201301].

\bibitem{Klein:2003qc}
  S.~Klein and J.~Nystrand,
  arXiv:hep-ph/0310223.

\bibitem{Baur:2003ar}
  G.~Baur, K.~Hencken, A.~Aste, D.~Trautmann and S.~R.~Klein,
  Nucl.\ Phys.\ A {\bf 729}, 787 (2003)
  [arXiv:nucl-th/0307031].

\bibitem{HenckenTBP05} K. Hencken, G. Baur, D. Trautmann, to be
  published (2005).

\bibitem{BaurB88}C.~Baur and C.~A.~Bertulani, Phys. Rep. {\bf 163},
  299 (1988).

\bibitem{BaurB93}
G. Baur and N. Baron, Nucl. Phys.~A {\bf 561},  628  (1993).

\bibitem{VidovicGB93}
M. Vidovic, M. Greiner, C. Best, and G. Soff, Phys. Rev.~C {\bf 47},  2308
  (1993).

\bibitem{Crittenden:1997yz}
  J.~A.~Crittenden,
  arXiv:hep-ex/9704009.

\bibitem{Kopeliovich:2001dz}
  B.~Z.~Kopeliovich, A.~V.~Tarasov and O.~O.~Voskresenskaya,
  Eur.\ Phys.\ J.\ A {\bf 11}, 345 (2001)
  [arXiv:hep-ph/0105110].

\bibitem{Frankfurt:2002wc}
  L.~Frankfurt, M.~Strikman and M.~Zhalov,
  Phys.\ Lett.\ B {\bf 537}, 51 (2002)
  [arXiv:hep-ph/0204175].

\bibitem{Nystrand:2002pd}
  J.~Nystrand, A.~J.~Baltz and S.~R.~Klein,
  arXiv:nucl-th/0203062.

\bibitem{Baur1991} 
G.Baur Nucl. Phys. A 531(1991)685

\end{thebibliography}
\end{document}